\documentclass[12pt]{iopart}

\usepackage{graphicx}
\begin{document}

\title[Loops, filaments and shock fronts in NGC~1275]{Turbulence and the formation of
filaments, loops and shock fronts in NGC~1275}

\author{D Falceta-Gon\c calves$^1$, E M de Gouveia Dal Pino$^2$, J S Gallagher$^3$
\& A Lazarian$^3$}

\address{1 - N\' ucleo de Astrof\'
isica Te\' orica, Universidade Cruzeiro do Sul - Rua Galv\~ ao Bueno
868, CEP 01506-000, S\~ao Paulo, Brazil; \\
2 - Instituto de Astronomia, Geof\'\i sica e Ci\^encias Atmosf\'ericas,
Universidade de S\~ao Paulo, Rua do Mat\~ao 1226, CEP 05508-900,
S\~ao Paulo, Brazil; \\
3 - Astronomy Department, University of Wisconsin,
 Madison, 475 N. Charter St., WI 53711, USA}
\eads{\mailto{diego.goncalves@cruzeirodosul.edu.br},  \mailto{dalpino@astro.iag.usp.br},
\mailto{jsg@facstaff.wisc.edu}, \mailto{alazarian@sfacstaff.wisc.edu}}

\begin{abstract}
NGC~1275, the central galaxy in the Perseus cluster,
is the host of gigantic hot bipolar bubbles inflated by AGN
jets  observed in the radio as Perseus~A.
It presents a spectacular
$H{\alpha}$-emitting nebulosity surrounding NGC~1275, with
loops and filaments of gas extending to over 50 kpc. The origin
of the filaments is still unknown, but probably correlates with the
mechanism responsible for the giant buoyant bubbles.
We present  2.5 and 3-dimensional MHD simulations of the central
region of the cluster in which turbulent energy,
possibly triggered by  star formation and supernovae (SNe)
explosions is introduced.
The simulations
reveal that the turbulence injected by massive stars could be
responsible for the nearly isotropic distribution of filaments and
loops that drag magnetic fields upward as indicated by recent
observations.  Weak
shell-like shock fronts propagating into the ICM with velocities of
100-500 km/s are found, also resembling the observations. The isotropic
outflow momentum of the turbulence  slows  the infall of
the intracluster medium, thus limiting further starburst activity
in NGC~1275. As the turbulence is subsonic over most of the simulated volume, the
turbulent kinetic energy  is not efficiently converted into heat
and additional heating  is required to suppress the cooling
flow at the core of the cluster. Simulations combining
the MHD turbulence with the AGN outflow can reproduce the
temperature radial profile observed around NGC~1275. While the AGN
mechanism is the main heating source, the supernovae are crucial to
$isotropize$ the  energy distribution.

\end{abstract}

\maketitle

\section{Introduction}

Perseus (Abell 426) at z = 0.0176 is the brightest galaxy
cluster in X-ray, is a $prototype$ of cooling core
clusters. Its central galaxy, NGC 1275, hosts a
narrow-line radio source, Per A (3C84), which interacts with the
intracluster gas through its jets and bipolar outflows.
{\it Chandra} X-ray observations (Fabian et al. 2003, 2006, Saunders \& Fabian 2007)
reveal  an associated complex intracluster medium (ICM),
with temperatures ranging from 2.5 keV near NGC~1275
up to 7 keV in the outskirts. Features like X-ray
cavities and spherical shock waves (Graham, Fabian \& Sanders 2008) observed in its
thermal plasma seem to be associated with  energy flows in the central
region.

A rich filamentary structure
of cold gas is roughly isotropically distributed
around NGC~1275 ($\sim 1-10^3$ cm$^{-3}$ and $T \sim 10^3 -
10^4$ K)(Conselice, Gallagher \& Wise 2001, Salome et al. 2006,
Fabian et al. 2003, Fabian et al. 2008). It extends to  $\sim
75$ kpc in radius and is $>10^8$ yr old, and its origin is still
undetermined (Johnstone et al 2007, Ferland et al. 2008, Revaz et al. 2008). The
filaments have radial velocities of  $\sim 200$ km s$^{-1}$, much
smaller than the sound speed of $\sim 700$ km s$^{-1}$
in the surrounding  hot ICM  (Hatch et al. 2006, Gallagher et al. in preparation).
The mechanism by which the filaments are stabilized against
disruption into the pervasive $4 \times 10^7$ K ICM is also
unclear, but a magnetic field in the filaments ($\sim 100 \mu$G) in
equipartition with the surrounding pressure would be sufficient to
assure their stabilization (Fabian et al. 2008).

It has been suggested that the formation of filaments and the increase of 
turbulent energy in galaxy clusters could be related to the cooling flows and the 
consequent magnetic compression (Pistinner \& Shaviv 1995, 
Godon et al. 1998). However, there is no evidence of strong cooling flows that could 
reproduce these models. Both the origins of the giant gas filaments and suppression 
of the cooling flow in A426 can be associated with the presence of giant bubbles of hot
gas inflated by the AGN (Fabian et al. 2003). However, recent
hydrodynamical simulations have shown that the AGN  feedback
seems to be insufficient to reduce the cooling flow effects to the
observed values, $T_{\rm ICM}/T_{\rm core} \sim 3$ (Gardini 2007).
Simulations also suggest that AGN are likely to be ineffective in
creating a nearly isotropic distribution of filaments. This is
because, despite the large AGN power output ($10^{42} - 10^{44}$ erg
s$^{-1}$), which explains the production of the hot bubbles,
the low density and momentum associated with the relativistic jets do
not readily distribute the thermal energy isotropically.
Additional energy/momentum mechanism(s) may be required to
produce both the isotropic filamentary structure and the
isotropization of energy.

The issue of AGN feedback in
galaxy clusters presents a number of unsolved problems yet which
have been extensively discussed in the literature. Besides those
discussed above, hydrodynamical simulations have also shown that the
AGN bubbles may disrupt within 100 Myr due to Kelvin-Helmholtz and
Rayleigh-Taylor instabilities (Bruggen et al. 2005, Heinz et
al.2006, Pizzolato \& Soker 2006), failing to reproduce the observed
ICM cavities, whose inferred ages are $\geq 10^8$ yrs in the outer
regions of Perseus (Nulsen et al. 2005). The inclusion of magnetic
fields in bubbles which are inflated by kinetic-dominated jets seems
to alleviate this problem (Robinson et al. 2004, Ruszkowski et al.
2007), but  also reduce the extent to which the interior of the
hot bubbles couple to the surrounding medium, making it much more
difficult for AGN heating to balance cooling (e.g., Bruggen et al.
2009). Even when the bubbles are
 inflated by magnetically-dominated jets (Nakamura et
al. 2007, Liu et al. 2008), they rise through
the cluster ICM as bipolar structures, and it is unclear how they
couple with the surrounding medium and distribute heating.

In this work, we discuss another aspect of these questions
through an examination of a supplementary stellar source of energy and
momentum injection. We assume that gas infall
into a central cluster galaxy triggers star formation and consequently SNe
explosions that produce turbulence.
SNe-driven turbulence in galaxy haloes is quite a common
phenomenon (see de Avillez 2000;  de Gouveia Dal Pino et al.
2009 for reviews). Edge-on star forming galaxies often exhibit hot
halos with structures that resemble chimneys and fountains extending
for several kpc above the galaxy. Numerical simulations
indicate that they are produced by SNe, which blow superbubbles that
carve the disk material and propels gas outwards (Melioli et al.
2008; 2009 and references therein). Halo cloud complexes and
filaments resulting from these processes
are observed in star forming disk galaxies
(e.g. Dettmar 2005). The role of SN-powered galactic winds
in preventing central gas cooling
 has been also recently investigated in
the context of fossil group of galaxies (Dupke et al. 2009).

This study explores the role of turbulence injected in the
central region of NGC~1275. Our model examines the importance
of turbulence in the suppressing  the cooling flow,
 providing energy/momentum for the production of
the complex filamentary structures of cold gas, and
dragging  magnetic energy outwards into the ICM.
To this end we made  magnetohydrodynamical (MHD)
numerical simulations of turbulence
evolution in a realistic distribution of  cluster gas, threaded by a
weak magnetic field. The model and initial setup of the simulations
are described in the following section. The main results and
discussions are presented in \S 3, and we draw our
conclusions in \S4.

\section{The model}

In order to simulate the role of turbulence on mass and energy
feedback in a galaxy cluster core, we performed
MHD simulations accounting for the hot gas
surrounding the core galaxy, and introduced the turbulent energy
injection at the central core that could represent the
feedback from SNe in an ongoing starburst.
The model was implemented in a
well-tested Godunov-MHD scheme, in which we integrate the full set
of MHD equations in conservative form (Falceta-Gon\c calves,
Kowal \& Lazarian 2008, Burkhart et al. 2009, Kowal et al. 2009, Le\~ao et al. 
2009).

An external force term, ${\bf f} = {\bf f_{\rm turb}}+{\bf f_{\rm
grav}}$, which is responsible for the turbulence injection and
gravity, is explicitly incorporated into the momentum equation.
The turbulence is introduced by a random solenoidal function in
Fourier space within a chosen range of scales (see Alvelius 1999 for details). The
effects of radiative cooling are treated separately, as we compute
$\frac{\partial P}{\partial t} = (1-\gamma) n^2 \Lambda(T)$, after
each timestep, where $n$ is the number density and $\Lambda(T)$ is
the interpolation function from an electron cooling efficiency table
for an optically thin gas (Gnat \& Sternberg 2007).

The external gravity is introduced through an
appropriately scaled fixed distribution of dark
matter following the NFW profile (Navarro, Frenk \& White 1996),
\begin{equation}
\rho_{\rm DM}(r)=\frac{\rho_s}{(r/r_s)(1+r/r_s)^2},
\end{equation}
\noindent where $r_s$ represents the characteristic radius of the
cluster and $\rho_s = M_s/(4\pi r_s^3)$. From isothermal pressure
equilibrium the gas density may be described by $\rho(r)=\rho_0
[\cosh(r/r_s)]^{-1}$.
As initial setup for the simulations we used an initial core density $\rho_0=5 \times
10^{-2}$cm$^{-3}$ and $r_s = 30$ kpc, which results in a similar
profile to the empirical density distribution (Sanders et al. 2004). The core temperature
is set as $T_0 = 7 \times 10^7$K which gives a free-free cooling timescale
of $\sim 400$Myr. The turbulence is introduced within the range
$1 <l_{\rm inj} < 3$ kpc. This range of values is a little large - due to limited 
resolution - whencompared to the typical sizes of galactic superbubbles
inflated by starbursts, which are of the order of several hundreds of parsecs. The injection
scales, however, do not play a major influence on the subject of this study. The 
turbulent energy injection occurs within a radius of 5 kpc around the core center, and is
set at a constant rate of $P_{\rm inj} = 10^{56}$erg Myr$^{-1}$. This value
represents injection from  $10^{-1}$ SNe/yr, in agreement with the
expected value for NGC~1275 where the star formation rate
is $\sim 30$ M$_{\odot}~yr^{-1}$ (Dixon et al. 1996).

Faraday rotation and synchrotron measurements obtained for several galaxy clusters 
have suggested magnetic field intensities of $B \sim 0.1 - 1 \mu$G, for the cluster 
halo. In our simulations, the magnetic field is assumed to be initially uniform in the 
x-direction, with intensity $B_0= 1 \mu$G, very small compared to 
the thermal pressure ($\beta\sim 10^4$) but in agreement with the observations. Because 
of the high beta value the initial topology 
of the field is dynamically irrelevant. The computational domain corresponds to a box
with physical size of $L = 100$ kpc. The cube is homogeneously divided into fixed 256$^3$
cells, corresponding to 0.39 kpc/cell. We have also used open
boundary conditions in order to allow gas motions in/outward the
computational domain as the pressure gradients evolve with time.

\section{Results}

\begin{figure*}[tbh]
\centering
\includegraphics[scale=.2]{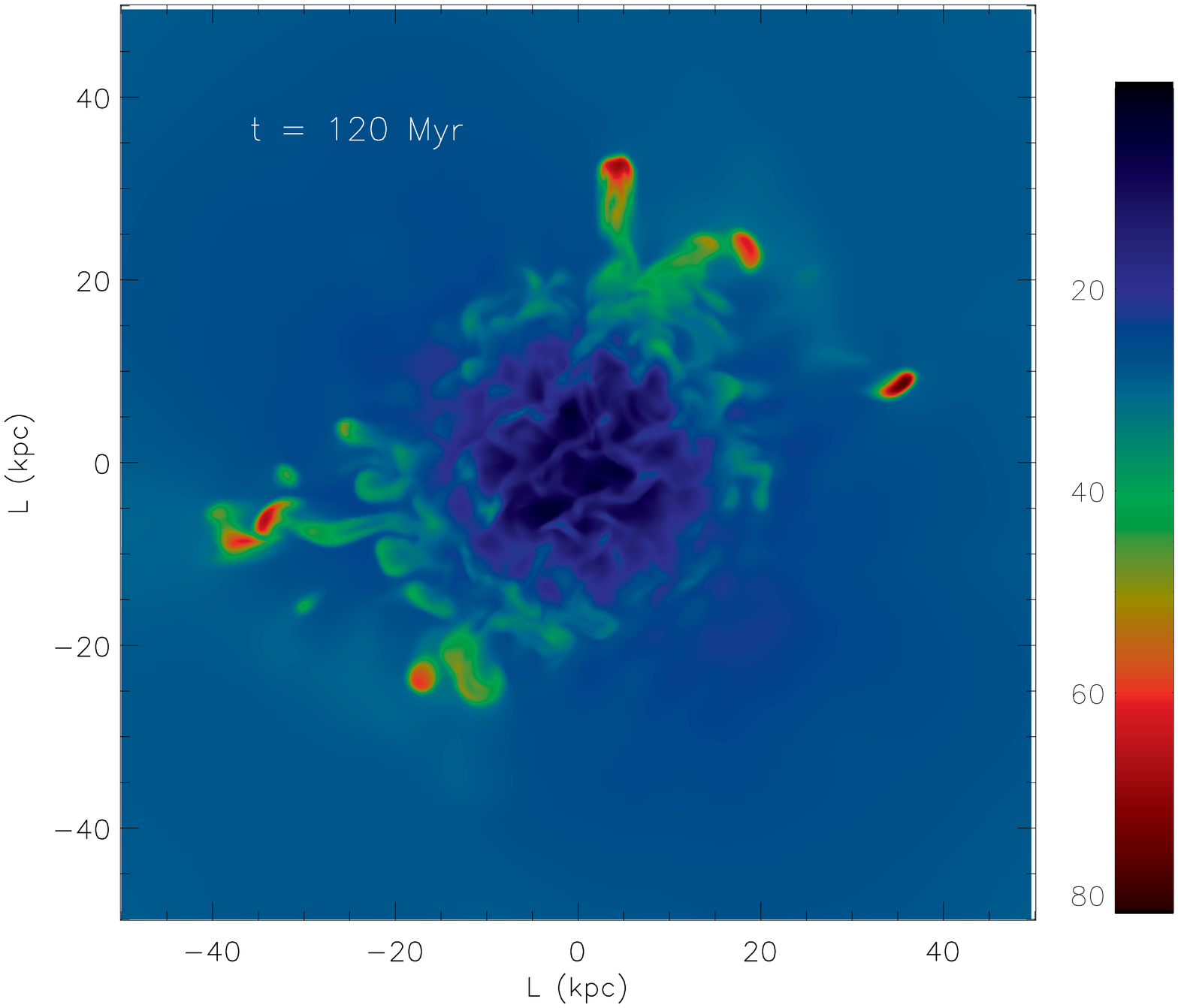}
\includegraphics[scale=.2]{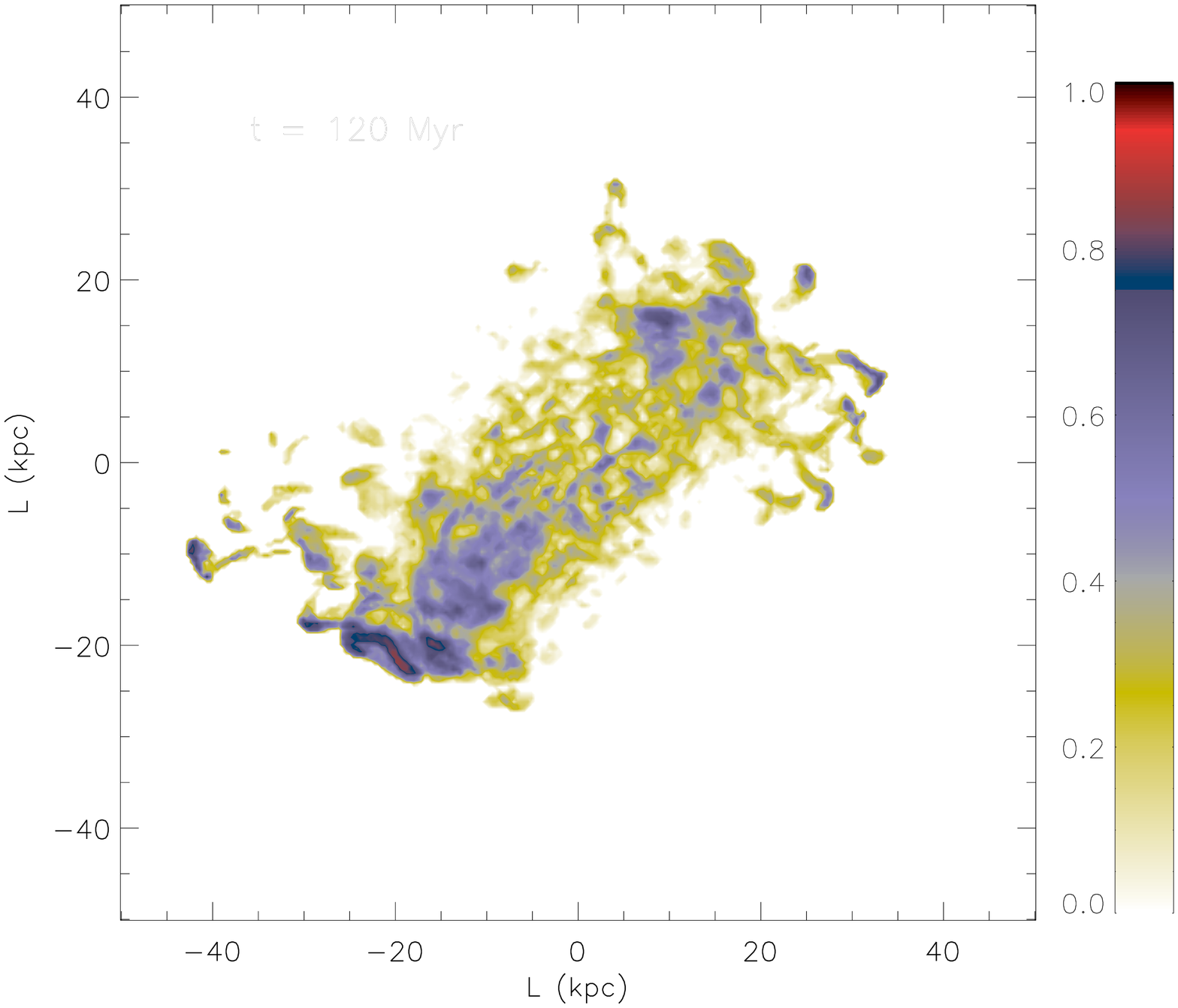}
\includegraphics[scale=.2]{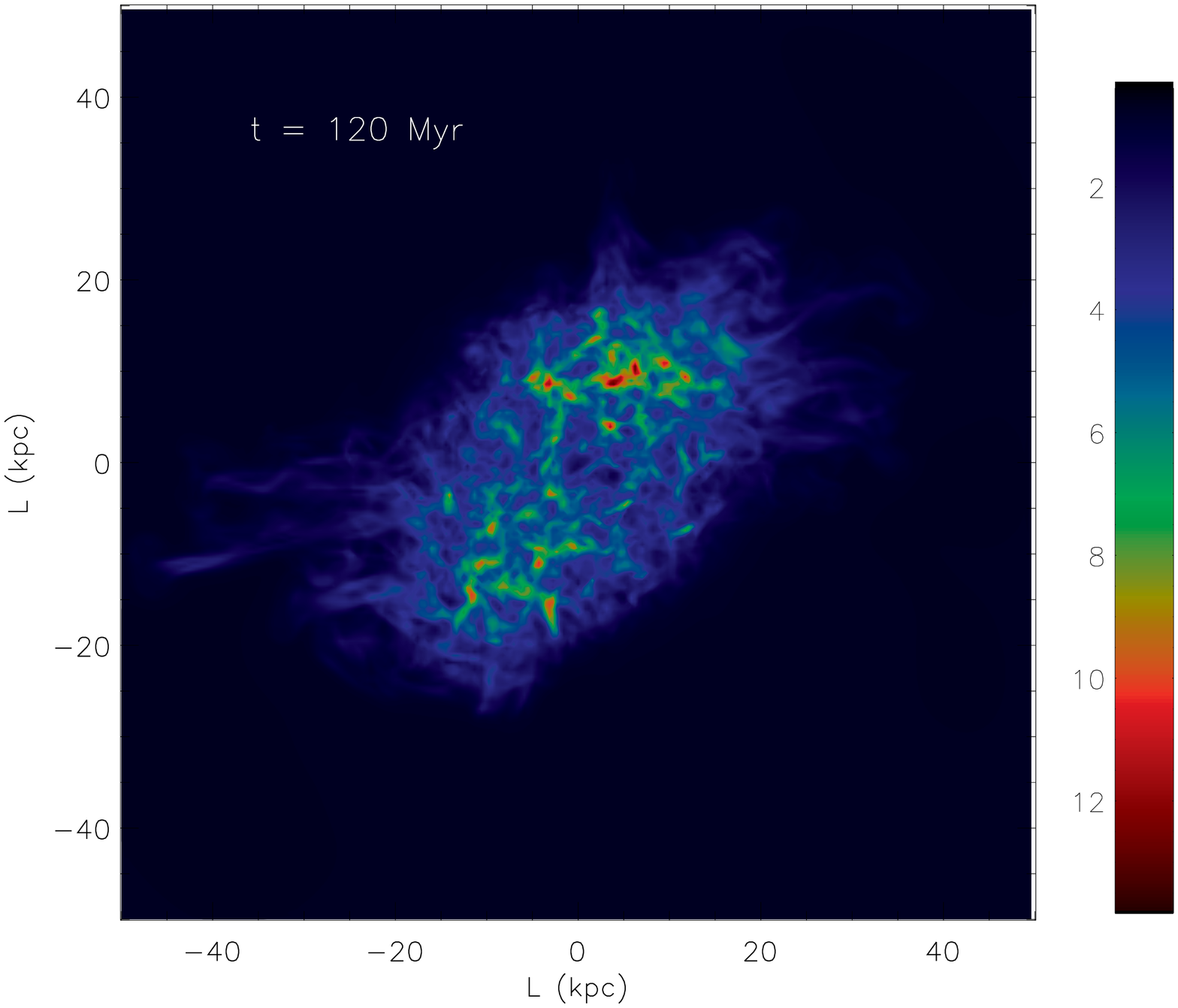}
\caption{$Left$: 2D slice of map of density normalized by local density at t=0.
$Center$: Emission map for the
overdense filamentary structures normalized by a maximum value (see
text). $Right$:
projection of magnetic pressure ratio ($\int_{LOS}(B^2/B_0^2) dl/L$).}
\end{figure*}

The simulations were carried out over a time span of
$t=150$ Myr, consistent with the extended period of star formation in NGC1275 based on 
ages of its massive star clusters (e.g. Carlson et al. 1998). Turbulence is 
initially responsible for small (low 
density) bubbles. At thesame time, as the gas starts to cool, there is an inward flux
as a cooling flow. At this stage, the small
cavities do not expand much since the shock heating is efficient only in the core region
where the sound speeds are small. As the distance to the center of the
system increases the sound speed increases and shock dissipation becomes less efficient.
However, as more turbulent cells are randomly injected,
eventually coinciding with previously inflated cavities,
further expansion is observed.

The interaction between the
SN-driven turbulence and the overdense structures results in the
formation of the gaseous filamentary structures, as shown in Figure 1 (left) at
$t=120$ Myr. The map of overdense structures was obtained by
dividing the density map of the central slice of the cube at $t=120$ Myr by its value at
$t=0$ Myr. In Figure 1 (center) we show the emission map for a given line of sight,
assuming the gas is optically thin. For this
plot we considered the emission (which is proportional do the density squared) from the
densest regions, above an arbitrary  threshold taken as 5 times the averaged density. The
contribution from cells below this threshold was set to zero. This calculation mimics the
optical emission from molecular gas, which occurs only at the densest regions. The 
filaments have sizes of $\leq$ 40 kpc, and move outward with average velocities of $\sim
100 - 500$ km s$^{-1}$. At the end, ram pressure exerted by the outward turbulent motion
counter-balances the incoming cooling flow; matter infall ceases at $t \sim 10$ Myr.

During the formation of
the filaments, the magnetic field is dragged and compressed with the
gas. The local density within the filaments is $n\sim 0.01 - 0.04$
cm$^{-3}$, a factor of $\sim 10 - 100$ times larger than the ambient value. As a 
result, the magnetic pressure within the overdense filaments and loops is typically 
larger thanthe surroundings by a factor of $30 - 200$, i.e. the filaments present
absolute values of magnetic field intensities in the range $B_{\rm
fil} = 5 - 20 \mu$G (as seen in Fig. 1 [right]).  
This value is comparable to the estimation of $B \sim 24 \mu$G (Fabian et al.
2008) based on stability conditions for gravitational support. The surroundings, on the 
other hand, present field intensities in the range $0.1 - 1 \mu$G, in agreement with 
synchrotron and farady rotation estimates (Carilli \& Taylor 2002). A strong magnetic 
field would also be responsible for a decrease 
in gas diffusion and thermal conductivity, which would allow the filaments to further 
decrease in size and increase in density due to cooling (Revaz et al. 2008). Higher gas 
densities in the filaments are required to match the observations.

Another important feature observed in the simulation is the
generation and propagation of acoustic waves. Fabian et al. (2003)
studied the excess emission in X-rays from Chandra observations and
identified weak shock fronts, or acoustic waves, propagating
outwards the inner part of Perseus. Figure 2 presents the two
dimensional line integration convolution method (LIC) applied to the
velocity field of the middle slice of the simulated cube. The
texture of the plot represents the orientation of the velocity
vectors, and the gray-scales represent its local intensity relative to the sound
speed (bright regions in the center show the turbulence injection scales). The
shell-like gradients in gray-scales represent the acoustic
waves, or even shocks, if present, propagating outwards as seen in
the observations. Fabian et al. (2003) have speculated about their
origin considering the effects of rising bubbles inflated by the AGN. The
isotropic distribution of the ripples and weak shocks indicates that they may be well explained by the turbulence itself.

\begin{figure}
\centering
\includegraphics[scale=.24]{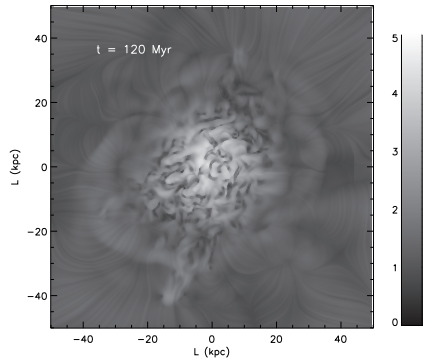}
\caption{Line integral convolution map for the velocity field using the
same time step as in Fig. 1 (t= 120 Myr). Gray-scales
correspond to velocity amplitudes which are related to the local sound speed (Mach 
number) and
texture corresponds to vectors orientation. Spherical shaped shock fronts propagate
outwards in different directions.}
\end{figure}

In spite of allowing the formation of loops and dense filamentary
structures, as well as the increase of magnetic field intensity
within ripples and filaments, and the generation of nearly isotropic
shock/acoustic waves, turbulence alone cannot be responsible for the
heating of the central region of the cluster. Even though ICM infall
 ceases, mostly due to the isotropic momentum distribution of the turbulent motions,
cooling  still dominates. Being subsonic over most of the
simulated volume, the turbulent kinetic energy is not efficiently
converted into heat. Therefore, another heating source is
required.

Actually, the AGN, although not providing much
momentum to the gas, is the main source of energy in NGC~1275 (and
possibly in most galaxy clusters). In order to test its importance
in heating the cluster core, we performed a 2.5-D simulation
with similar setup as described in the run above, but including an
AGN ``heavy" particles jet, i.e. of ions, with a total power of $L = 10^{43}$ erg 
s$^{-1}$.
For that we use a standard procedure ($see$ Heinz et al. 2006), selecting 2 cells in the
center of the simulated domain and fixing the local
velocity as $v_{\rm jet} = 20 c_s$ (which corresponds to $\sim 10^4$ km$s^{-1}$) in both
opposite directions along the x-axis. The jet temperature is set as 
$T_{\rm jet} = 10 T_0$, and its density as $n_{\rm jet} = 0.1 n_0$,
resulting in a total mass loss rate of $\sim 0.2 M_{\odot} yr^{-1}$.

In Figure 3, we
compare the average radial temperature profiles obtained for each model, i.e. with and
without the inclusion of the AGN. The AGN represents the main
source of heating in the system, but turbulence is still required to
$isotropize$ its distribution. It is important to remark that,
though we do not show the maps here, the density structures, as
observed in the slices of the 3-dimensional MHD turbulence discussed
above, are almost the same in the model including the AGN. This follows
because the AGN momentum is much smaller than that injected by the
SN-driven turbulence.

\begin{figure}
\centering
\includegraphics[scale=.3]{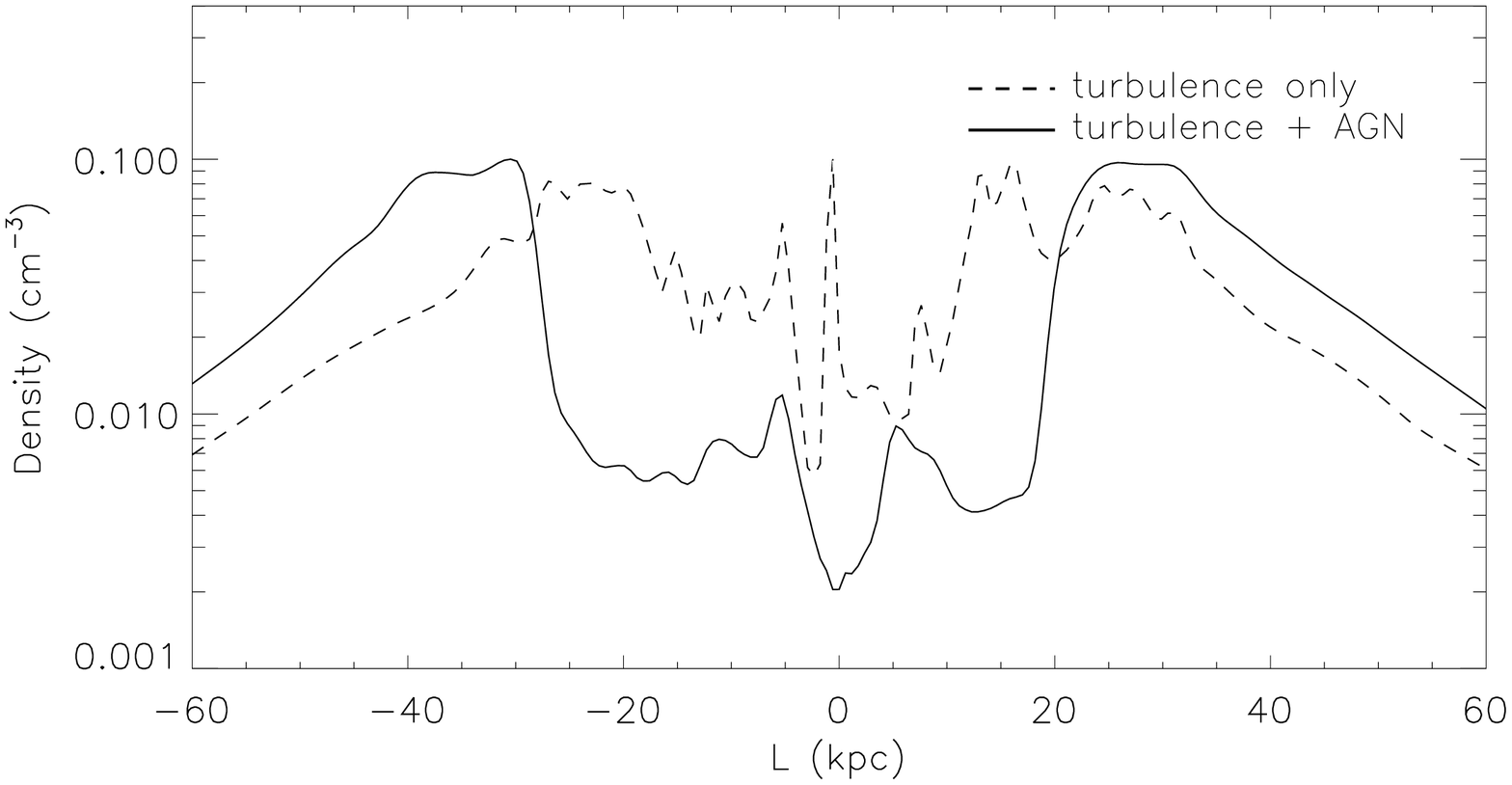} \\
\includegraphics[scale=.3]{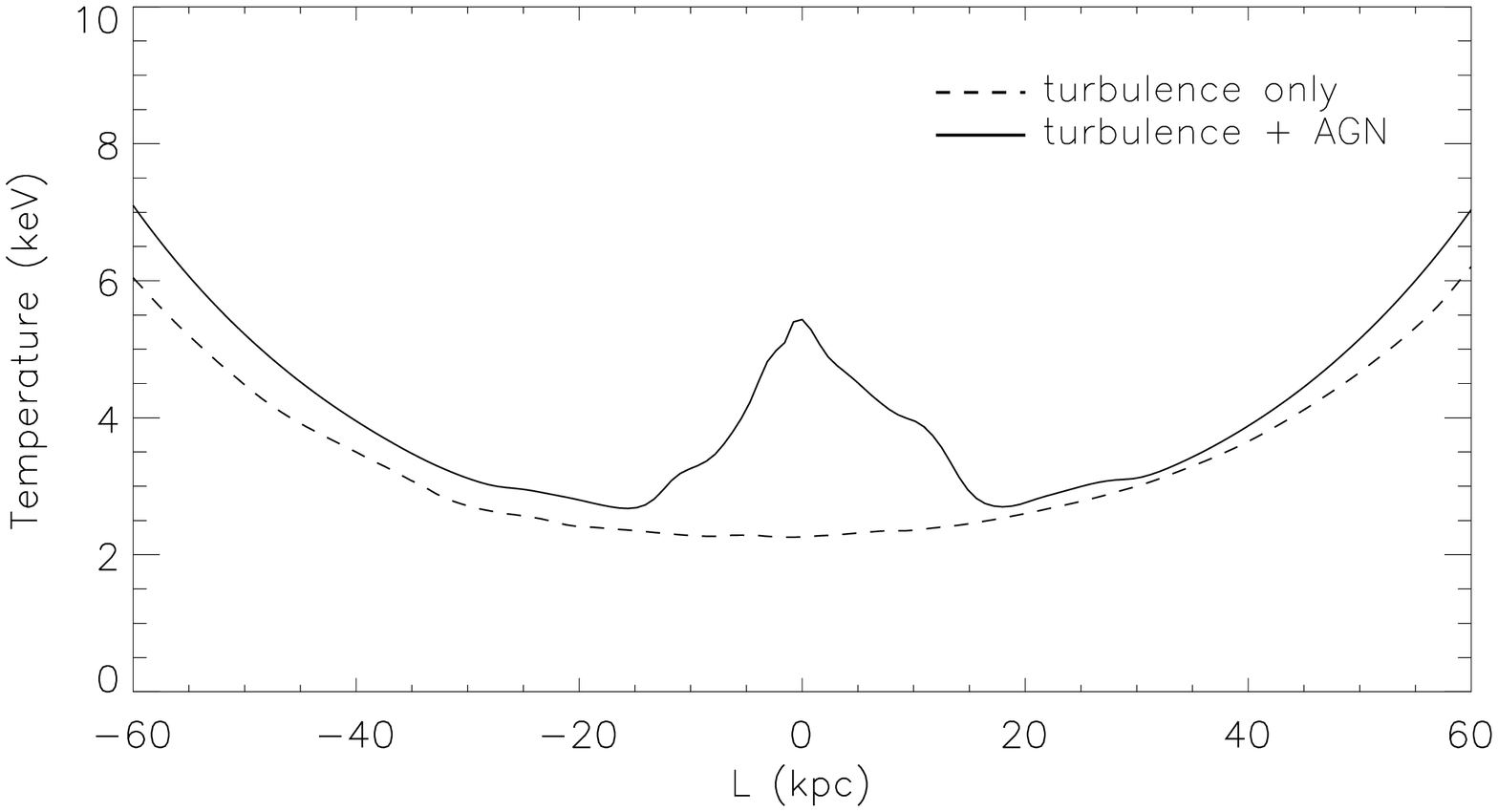}
\caption{Density (top) and temperature (bottom) profiles as
obtained for a 2.5 D turbulent model without AGN heating
(solid line) and turbulence plus AGN heating model (dashed line).}
\end{figure}

\section{Conclusions}

In this work we presented a study of the role of MHD turbulence in
the formation of large scale density structures in the central
regions of galaxy clusters with the objective of better understanding
the giant gas filaments in NGC~1275.   Since AGNs do not
seem to provide enough momentum to distribute their power
isotropically around the cluster core, they may not be the
main mechanisms responsible for filamentary and shell-like ISM structures
found, e.g., in the Perseus galaxy cluster. Our model provides
additional energy/momentum to the intracluster plasma
via SNe explosions which we relate to the high star formation rate
in NGC~1275.  These can produce
shocks that propagate outwards, pushing enough
of the ISM to qualitatively reproduce the observations.

Our numerical simulations of MHD turbulence injected in a realistic
ICM have shown that most of the peculiar structures observed in
NGC~1275 might be reproduced if a SNe injection power of $10^{56}$ erg
Myr$^{-1}$ (typical for starburst galaxies) is used. This process
raises gas loops and filaments  with average velocities of
$\sim 100 - 500$ km s$^{-1}$, compatible with the observations. Some filaments are
 related to bubble motions. This is not surprising as the turbulence itself results in
the creation of small scaled bubbles that evolve being dragged outwards. Compared to AGN
inflated bubble models, though, our model results in more isotropic distributions and
presents shorter evolutionary timescales. These structures are also threaded
by magnetic fields whose intensity is amplified by compression. In order to verify 
this possibility we have calculated the {\it B versus n} correlation. For the filaments 
we obtained $B \propto n^{0.87}$. The power index, close to unity, is expected for 
magnetic compression at shocks. Magnetic fluctuations due to MHD waves, on the 
other hand, result in power index close to 0.5, which is comparable to the value 0.41 
obtained for the halo, disregarding the filaments. The resulting magnetic pressure is 
similar to the required values of $\sim 25 \mu$G 
to help stabilize them against gravitational collapse and diffusion, while the magnetic 
field at the halo is kept in the range $0.1 - 1 \mu$G. The major role of the magnetic 
field is to stabilize the filamentary structures.
For example, magnetic field reduced the effect of Kelvin-Helmholtz instability at the
edge of the structures and helped preventing the diffusion of the cooled material with the
surroundings. Of course, the transport issue is quite important to address (see e.g.,
Parrish, Quataert \& Sharma 2009), though it has not been calculated in our models.
However, as pointed by Dennis \& Chandran (2005), the energy transfer rate by heat
conduction is very small compared to the dynamical sources.
Because of the isotropic momentum injection, the ICM infall due to
the cooling flow is ceased at $t \sim 10$ Myr. However, the
temperature at the core is not increased, as expected from the
observations. This is an indication that turbulence may be the main
source of momentum to the system, but not the main heating source.
From  2.5-D simulations of MHD turbulence working together with a
typical AGN of $L = 10^{42}$ erg s$^{-1}$, we also reproduced
the radial temperature profile observed in NGC~1275.

Our results have special importance for
understanding the halting of cooling flows in  galaxy
clusters. AGN may provide enough energy for the heating of the
cooled gas in the cluster cores, but turbulence - even though not
necessarily related to starbursts - is needed to isotropize and
distribute the energy. A more detailed analysis, including fully
3-dimensional modeling also with AGN heating, as well as models with
finer resolution, is currently in process and will be presented in a
future work.

\medskip

The authors acknowledge support from
FAPESP and CNPq grants.  JSG received support from NSF grant AST-0708967  to the University of Wisconsin.


\medskip

\begin{thebibliography}{}


\bibitem{alvelius99} Alvelius K. 1999, Phys. Fluids, 11, 7, 1880


\bibitem{bruggen05} Bruggen M., Ruszkowski
M. \& Hallman E. 2005, ApJ, 630, 740

\bibitem{bruggen09} Bruggen M., Scannapieco E. \&
Heinz S. 2009, MNRAS, 395, 2210


\bibitem{burkhart09} Burkhart, B.; Falceta-Gonçalves, D., Kowal, G., Lazarian, A. 2009, 
ApJ, 693, 250

\bibitem{carilli02} Carilli C. L. \& Taylor, G. B. 2002, ARA\&A, 40, 319

\bibitem{carl98} Carlson M. N. et al. 1998, AJ, 115, 1778

\bibitem{cons01} Conselice C. J., Gallagher J. S. \&
Wyse R. F. G. 2001, ApJ, 211, 135

\bibitem{avil00} de Avillez, M. 2000, MNRAS, 315, 479

\bibitem{gdp2009}
de Gouveia Dal Pino E. M., Melioli C., D'Ercole A., Brighenti
F. C. \& Raga A. 2009, Adv. Space Res., in press

\bibitem{dennis05}
Dennis T. J. \& Chandran B. D. G. 2005, ApJ, 622, 205

\bibitem{detsoida2009}
Dettmar R.-J. \& Soida M. 2006, Ast. Nachrichten, 327, 495

\bibitem{1996AJ....111..130D} Dixon, W.~V.~D.,
Davidsen, A.~F., \& Ferguson, H.~C.\ 1996, AJ, 111, 130

\bibitem{dupke2009}
Dupke R., Mendes de Oliveira C. \& Sodr\'e L. 2009, ApJ (submitted)

\bibitem{fabian03} Fabian, A.C., Sanders J. S., Allen
S. W. et al. 2003, MNRAS, 344, 43

\bibitem{fabian06} Fabian, A.C., Sanders J. S., Taylor
G. B. et al. 2006, MNRAS, 366, 417

\bibitem{fabian08} Fabian A. C., Johnstone R. M., Sanders J. S. et
al. 2008, Nature, 454, 968

\bibitem{falceta08} Falceta-Gon\c calves
D., Lazarian A. \& Kowal G. 2008, ApJ, 679, 537


\bibitem{ferland08} Ferland, G. J., Fabian, A. C., Hatch, N. A. et
al. 2008, MNRAS, 386, 72

\bibitem{gardini07} Gardini A. 2007, A\&A, 464, 143

\bibitem{gnat07} Gnat O. \& Sternberg A. 2007, ApJ, 168, 213

\bibitem{1998AJ....116...37G} Godon, P., Soker, N.,
\& White, R.~E., III 1998, AJ, 116, 37

\bibitem{graham08} Graham J., Fabian A. C. \& Sanders J.
S. 2008, MNRAS, 386, 278

\bibitem{2006MNRAS.367..433H} Hatch, N.~A., Crawford,
C.~S., Johnstone, R.~M., \& Fabian, A.~C.\ 2006, MNRAS, 367, 433


\bibitem{heiz06} Heinz S., Bruggen M., Young A. \& Levesque E. 2006,
MNRAS, 373, 65

\bibitem{john07} Johnstone R. M., Hatch N. A., Ferland G.
J. et al. 2007, MNRAS, 382, 1246


\bibitem{kowal09} Kowal, G., Lazarian A., Vishniac E., \& Otmianowska-Mazur, K. 2009, ApJ, 700, 63

\bibitem{leao09} Le\~ao, M. R. M., de Gouveia Dal Pino, E. M., Falceta-Gon\c calves, D.,
Melioli, C., Geraissate, F. 2009, MNRAS, 394, 157


\bibitem{liu08} Liu W., Li H., Li S. \& Hau S. 2008, ApJ, 684, 57

\bibitem{mel2008}
Melioli C., Brighenti, F. C., D'Ercole, A. \& de Gouveia Dal Pino, E. M. 2008, MNRAS,
388, 573

\bibitem{mel2009}
Melioli C., Brighenti, F. C., D'Ercole, A. \& de Gouveia Dal Pino, E. M. 2009, MNRAS, 
399, 1089

\bibitem{naka07} Nakamura M., Li H. \& Li S. 2007, ApJ, 656,
721

\bibitem{nfw96} Navarro J. F., Frenk C. S. \& White S. D.
M. 1996, ApJ, 462, 563

\bibitem{nul05} Nulsen P. E. J., Hambrick D. C., McNamara
B. R., et al. 2005, ApJ, 625, 9

\bibitem{par09} Parrish I. J., Quataert E. \& Sharma P. 2009, ApJ,
703, 96

\bibitem{1995ApJ...446L..11P} Pistinner, S., \& Shaviv, G.\
1995, ApJL, 446, L11

\bibitem{piz06} Pizzolato F., \& Soker N. 2006, MNRAS,
371, 1835

\bibitem{rev04} Revaz Y., Combes F. \& Salom\'{e} P. 2008, A\&A, 477, 33

\bibitem{rob04} Robinson K., Dursi L. J., Ricker P. M. et al.
2004, ApJ, 601, 621

\bibitem{rus07} Ruszkowski M., Ensslin T. A., Bruggen M., et al.
2007, MNRAS, 378, 662

\bibitem{salome06} Salom\'e P., Combes F., Edge A. C. et al. 2006,
A\&A, 454, 437

\bibitem{sanders04} Sanders J. S., Fabian A. C., Allen S. W. \&
Schmidt R. W. 2004, MNRAS, 349, 952

\bibitem{sanders07} Sanders J. S. \&  Fabian A. C. 2007, MNRAS,
381, 1381

\end{thebibliography}
\end{document}